\documentclass[10pt,journal,epsfig]{IEEEtran}

\usepackage[dvips]{graphicx}
\usepackage{graphicx}
\usepackage{amssymb}
\usepackage{cite}
\usepackage{subfigure}
\usepackage{amsmath}
\usepackage{algorithm}
\usepackage{algorithmic}
\usepackage{multirow}

\begin{document}

\title{Joint Active and Passive Beamforming for Intelligent Reflecting Surface-Assisted Massive MIMO Systems}

\author{Xingjian Li, Jun Fang, Feifei Gao, and Hongbin Li
\thanks{Xingjian Li, and Jun Fang are with the National Key Laboratory
of Science and Technology on Communications, University of
Electronic Science and Technology of China, Chengdu 611731, China,
Email: JunFang@uestc.edu.cn}
\thanks{Feifei Gao is with the
Institute of Information Processing, Department of Automation,
Tsinghua University, Beijing 100084, China, Email:
feifeigao@tsinghua.edu.cn}
\thanks{Hongbin Li is
with the Department of Electrical and Computer Engineering,
Stevens Institute of Technology, Hoboken, NJ 07030, USA, E-mail:
Hongbin.Li@stevens.edu}
\thanks{This work was supported in part by the National Science
Foundation of China under Grant 61829103.}}

\maketitle

\begin{abstract}
In this paper, we study the problem of joint active and passive
beamforming for intelligent reflecting surface (IRS)-assisted
massive MIMO systems, where multiple IRSs equipped with a large
number of passive elements are deployed to assist a base station
(BS) to simultaneously serve a small number of single-antenna
users in the same time-frequency resource. Our objective is to
maximize the minimum signal to interference plus noise (SINR) at
users by jointly optimizing the transmit precoding vector at the
BS and phase shift parameters at IRSs. We show that an interesting
automatic interference cancelation (AIC) property holds
asymptotically as the number of passive elements approaches
infinity, i.e., when an IRS is optimally tuned to serve a certain
user, this IRS will become interference-free to other users. By
utilizing this property, the max-min problem can be converted into
an IRS-user association problem, where the objective is to
determine which IRSs are assigned for each user. An exhaustive
search scheme and a greedy search scheme are proposed to solve the
IRS-user association problem. Our theoretical analysis reveals
that our proposed solution attains an SINR that scales
quadratically with the number of reflecting elements. Also, our
theoretical result suggests that even with a moderate number of
active antennas at the BS, a massive MIMO like gain can be
achieved by increasing the number of passive reflecting elements,
thus significantly reducing the energy consumption at the BS.
Simulation results are provided to corroborate our theoretical
results and to illustrate the effectiveness of our proposed
solution.
\end{abstract}


%By exploiting the asymptotic orthogonality among channel vectors
%associated with different users, massive MIMO systems can achieve
%almost perfect inter-user interference cancelation with a simple
%linear precoder and receive combiner, and thus have the potential
%to enhance the spectrum efficiency by orders of magnitude.

%For example, in mmWave systems, a very small obstacle, such as a
%person's arm, can even block the link because of the narrow
%beamwidth of mmWave signals [].

\section{Introduction}
Massive multiple-input multiple-output (MIMO) is a promising
technology to meet the ever growing demands for higher throughput
and better quality-of-service of the fifth-generation (5G) and
beyond wireless networks
\cite{RusekPersson13,LarssonEdfors14,Marzetta10}. However, the
high hardware complexity and cost required by massive MIMO systems
are still the main roadblock that hinders its implementation in
practice, especially in the high frequency band such as millimeter
wave (mmWave) bands \cite{BuzziI16,ZhangWu17}. Moreover, due to
unfavorable propagation conditions, the link between the base
station (BS) and users might be highly vulnerable to blockages,
thus making the communication unstable and inefficient
\cite{TanSun18,AbariBharadia17}. Therefore, it is necessary to
develop new spectrum and energy efficient technologies for massive
MIMO systems.

%to maintain desirable system performance.

%In addition, since the reflecting elements are usually very low
%cost and energy consumption units, the IRS is cost and energy
%efficient and is feasible to implement in practice [].

In order to achieve high beamforming gains with low-cost systems,
intelligent reflecting surface (IRS), also known as large
intelligent surface (LIS), has been proposed as a promising
technology in recent years
\cite{LiaskosNie18,HuRusek18,LiangLong19,WuZhang19d}. IRS is a
planar array made of newly developed metamaterial, consisting of a
large number of cost-effective and energy-efficient passive
reflecting elements \cite{WuZhang19d,HuangAlexandropoulos18}. Each
reflecting element is controlled by a smart micro controller so
that the incident signal can be reflected with reconfigurable
amplitudes and phase shifts. By adaptively tuning the phase shifts
of reflecting elements, IRSs are able to enhance the received
signal power or suppress the co-channel interference for desired
users, thus improving the coverage and performance of wireless
systems \cite{WuZhang19b}.

%where the problem of jointly optimizing the active beamforming at
%the transmitter and passive beamforming at the IRS is of the most
%concern

IRS-aided wireless communications have attracted much attention
recently
\cite{WuZhang18,WangFang19a,NadeemKammoun19,WuZhang19a,WuZhang19b,HanTang19,
WuZhang19c,WangFang19b,HeYuan19,HuangZappone18,YangZhang19,CuiZhang19,
YuXu19,ShenXu19,GuanWu19,MishraJohansson19}. In
\cite{WuZhang18,WangFang19a}, it was shown that for a single-user
scenario, the IRS-assisted system can obtain a received signal
power gain in the order of $\mathcal{O}(M^2)$, compared with the
conventional massive MIMO system that achieves a received signal
power gain in the order of $\mathcal{O}(N)$. Here $N$ denotes the
number of antennas at the transmitter and $M$ denotes the number
of reflecting elements at the IRS. Such an improvement is due to
the fact that IRS works as the receiver and transmitter
simultaneously. In \cite{NadeemKammoun19}, authors studied the
problem of maximizing the minimum
signal-to-interference-plus-noise ratio (SINR) in a multi-user
scenario. It was empirically shown that for the multi-user case,
IRS-assisted systems can offer massive MIMO like gains with a much
fewer number of active antennas. Most prior works on joint active
and passive beamforming, e.g.
\cite{WuZhang19a,WuZhang19b,WuZhang18,NadeemKammoun19,WangFang19a,HanTang19},
either focus on the scenario where only a single IRS is employed
or the scenario where multiple IRSs are deployed to serve a single
user. The scenario where multiple IRSs are used to serve multiple
users has been rarely investigated. The work \cite{WuZhang19c}
studied the joint beamforming problem in such a scenario, and
their aim is to minimize the total transmit power under an
individual SINR constraint for each user.

%A similar system setup was also considered in \cite{CaoLv19}, with
%the objective of maximizing the weighted sum of SINRs associated
%with different users.

%In [], the problem of minimizing the total transmit power at the
%transmitter was studied, subject to a number of quality-of-service
%(QoS) constraints, i.e. the SINR of each user exceeds a certain
%threshold.

%An important theoretical finding ,
%To address this challenging problem, a key insight

%for both the line-of-sight (LOS)-dominated IRS-user channel and
%the Rayleigh IRS-user channel. The AIC property is important as it
%can help avoid complicated inter-IRS interference management

In this paper, we consider the scenario where multiple IRSs
equipped with a large number of passive reflecting elements are
deployed to assist the BS to simultaneously serve a small number
of single-antenna users. Our objective is to maximize the minimum
SINR at users by jointly optimizing the transmit precoding vector
at the BS and phase shift parameters at IRSs. An important
theoretical finding made in this paper is that when the phase
shift parameters of an IRS are optimally tuned to serve a certain
user, due to the asymptotic orthogonality among channel vectors
associated with different users, this IRS will become
interference-free to other users. Such a property is referred to
as automatic interference cancelation (AIC), and is proved to hold
valid asymptotically when $M\rightarrow\infty$ for both the
line-of-sight (LOS)-dominated IRS-user channel and the Rayleigh
IRS-user channel. This property is important as it can help avoid
complicated inter-IRS interference management. By utilizing this
property, the max-min problem can be converted into an IRS-user
association problem whose objective is to determine which IRSs are
assigned for each user. Our theoretical analysis reveals that the
proposed solution attains an SINR in the order of $\mathcal{O}(M^2
N)$, i.e., it scales quadratically with the number of reflecting
elements. Also, this result suggests that even with a moderate
number of active antennas at the BS, increasing the number of
passive reflecting elements can help achieve massive MIMO like
gains, thus significantly reducing the energy consumption at the
BS.

%Two schemes, namely, an exhaustive search scheme and a greedy
%search scheme, are proposed to solve the user association problem.

The rest of the paper is organized as follows. In Section
\ref{sec:system-model}, the system model and the joint active and
passive beamforming problem are discussed. The proposed joint
active and passive beamforming method is provided in Section
\ref{sec:proposed-method}. The automatic interference cancelation
property is discussed and proved in Section \ref{sec:AIC}. The
theoretical analysis of our proposed solution is presented in
Section \ref{sec:theoretical-analysis}. Simulation results are
provided in Section \ref{sec:simulation-results}, followed by
concluding remarks in Section \ref{sec:conclusions}.

%In some emerging scenarios such as mmWave communications, the
%direct link between the BS and the user may not be available due
%to unfavorable propagation conditions.

%is very likely to hold valid

\section{System Model and Basic Assumptions} \label{sec:system-model}
Consider an IRS-assisted massive MIMO system, where multiple IRSs
are deployed to assist the BS equipped with $N$ antennas to
simultaneously serve $K$ ($N> K$) single-antenna users in the same
time-frequency resource. Suppose $L$ IRSs are employed, and the
number of reflecting units at each IRS is denoted by $M$ ($M\gg
K$). Let $\boldsymbol{G}_l \in \mathbb{C}^{M \times N}$ denote the
channel from the BS to the $l$th IRS, and $\boldsymbol{h}_{l,k}
\in \mathbb{C}^{M}$ denote the channel from the $l$th IRS to the
$k$th user. In this paper, we neglect the direct link from the BS
to each user in order to simplify our analysis and better
understand the impact of IRSs on the system performance. Also, in
this paper, we assume
\begin{itemize}
\item[A1] The channel between the BS and each IRS is line-of-sight
(LOS) dominated and has a rank-one structure.
\end{itemize}
This assumption is very likely to be met in practice because the
IRS is usually installed on the facade of a high rise building in
the vicinity of the BS \cite{NadeemKammoun19}. As a result, the
channel matrix between the BS and the IRS is dominated by the LOS
component. Particularly, for mmWave frequency bands, measurement
campaigns have shown that the power of LOS path is much higher
than the sum of power of NLOS paths
\cite{Muhi-EldeenIvrissimtzis10}. Thus the channel can be
well-approximated as a rank-one matrix. Specifically, the BS-IRS
channel $\boldsymbol{G}_l$ can be expressed as
\begin{align}
  \boldsymbol{G}_l = \sqrt{NM}\alpha_l \boldsymbol{a}_r(\varphi_l,\vartheta_l) \boldsymbol{a}_t^H(\psi_l)
  \label{Gl}
\end{align}
where $\alpha_l$ is the complex gain, $\varphi_l\in [0, 2\pi]$,
$\vartheta_l\in [0, 2\pi]$, and $\psi_l \in [0, 2\pi]$ are the
associated azimuth angle of arrival (AoA), elevation AoA, and
azimuth angle of departure (AoD) respectively, and
$\boldsymbol{a}_r\in\mathbb{C}^{M}$
($\boldsymbol{a}_t\in\mathbb{C}^{N}$) is the array response vector
associated with the IRS (BS). Due to the rank-one structure of the
BS-IRS channel, we need to assume
\begin{itemize}
\item[A2] The number of IRSs is no less than the number of
users, i.e. $L\geq K$.
\end{itemize}
Otherwise it is impossible for the BS to serve $K$ users
simultaneously. On the other hand, as a cost-effective and
energy-efficient technology, it is envisioned that, in future
wireless networks, IRSs are densely deployed to provide favorable
propagation environments. In addition, we assume
\begin{itemize}
\item[A3] The transmit array response vectors
$\{\boldsymbol{a}_t(\psi_l)\}$ are near orthogonal to each other,
i.e.
$|\boldsymbol{a}_t^H(\psi_{i})\boldsymbol{a}_t(\psi_{j})|\approx
0$ for $i\neq j$.
\end{itemize}
Such an assumption is reasonable because to enhance the signal
coverage, IRSs are usually deployed surrounding the BS and as a
result, the AoD parameters $\{\psi_l\}$ are expected to be
sufficiently separated from each other. Particularly, when a large
number of antennas are employed at the BS, the near-orthogonality
property holds valid even for slightly separated AoD parameters
$\{\psi_l\}$.

\begin{figure}[!t]
\centering
\includegraphics[width=3.5in]{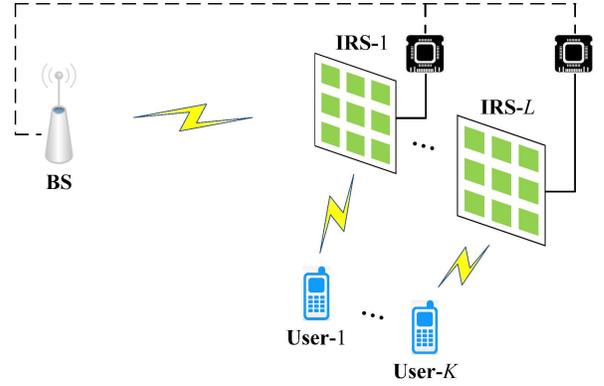}
\caption{Schematic of IRS-assisted massive MIMO systems.}
\label{System-model}
\end{figure}

%are sufficiently separated from each other.

%Second, to enhance the signal coverage, IRSs are usually deployed
%around the BS and as a result,

%due to the following practical considerations. First, in massive
%MIMO systems,

With the aid of a smart micro controller, each element of the IRS
can independently reflect the incident signal with a
reconfigurable phase shift \cite{WuZhang19a}. Therefore the
composite channel between the BS and the $k$th user is given by
\begin{align}
\boldsymbol{h}_k = \sum_{l=1}^L \boldsymbol{G}_l^H
\boldsymbol{\Phi}_l^H \boldsymbol{h}_{l,k}, \quad \forall k
  \label{hk}
\end{align}
where
\begin{align}
\boldsymbol{\Phi}_l\triangleq\text{diag}(
e^{j\theta_{l,1}},\ldots, e^{j\theta_{l,M}})
\end{align}
is the phase-shift matrix associated with the $l$th IRS, in which
$\theta_{l,m} \in [0,2\pi]$ denotes the phase shift associated
with the $m$th passive element of the $l$th IRS. The signal
received at the $k$th user can then be expressed as
\begin{align}
  y_k = \boldsymbol{h}_k^H \sum_{i=1}^K \sqrt{p_i}\boldsymbol{f}_i s_i + z_k
  \label{received-signal}
\end{align}
where $\boldsymbol{f}_i\in \mathbb{C}^{N}$ with
$\|\boldsymbol{f}_i\|_2=1$ is the transmit beamforming vector for
the $i$th user, $s_i\sim \mathcal{CN}(0,1)$ is the transmitted
symbol for the $i$th user which is assumed to be independent and
identically distributed (i.i.d.) random variables with unit
variance, $p_i$ is the transmit power allocated for the $i$th user
and we have the constraint $\sum_{i=1}^K p_i \leq P$, in which $P$
is the total transmit power, and $z_k$ denotes the additive
complex Gaussian noise with zero mean and variance $\sigma_z^2$.
Therefore, the signal to interference plus noise ratio (SINR) of
the $k$th user is given by
\begin{align}
  \text{SINR}_k = \frac{p_k\left|\boldsymbol{h}_k^H\boldsymbol{f}_k\right|^2}
  {\sum_{i\neq k} p_i\left|\boldsymbol{h}_k^H\boldsymbol{f}_i\right|^2 + \sigma_z^2}
\end{align}

Assuming the knowledge of the global channel state information,
our objective is to devise the transmit precoding vectors
$\{\boldsymbol{f}_k\}_{k=1}^K$, the transmit power allocation
$\{p_k\}_{k=1}^K$, and the phase shift matrices
$\{\boldsymbol{\Phi}_l\}_{l=1}^L$ to maximize the minimum SINR
among SINRs associated with all users. Such a problem can be
formulated as a max-min problem:
\begin{align}
  \max_{\{\boldsymbol{f}_k\},\{p_k\},\{\boldsymbol{\Phi}_l\}}\min_k\quad &\text{SINR}_k \nonumber \\
  \text{s.t.}\quad &\left\|\boldsymbol{f}_k\right\|_2=1, \forall k \nonumber \\
  & \theta_{l,m}\in[0,2\pi), \forall l, \forall m \nonumber \\
  & \sum_{k=1}^K p_k \leq P
  \label{Optimization-P1}
\end{align}
In the following, we will show that IRS equipped with a large
number of passive elements enjoys an appealing automatic
interference cancelation property. By resorting to such a
property, the above max-min problem can be converted into an
IRS-user association problem, which can be analytically solved.

%we develop an analytical solution to (\ref{Optimization-P1}).

\section{Proposed Method} \label{sec:proposed-method}
Our proposed method can be divided into two steps. In the first
step, given passive beamforming parameters
$\{\theta_{l,m}\}_{l,m}$, we optimize active beamforming
parameters, i.e. the transmit precoding vectors
$\{\boldsymbol{f}_k\}_{k=1}^K$ and the transmit power
$\{p_k\}_{k=1}^K$. The optimal active beamforming parameters are
then substituted into (\ref{Optimization-P1}) to obtain a passive
beamforming problem. In the second step, we proposed an effective
and analytical solution to this passive beamforming problem.

%Then, we search for the best precoding vectors
%$\{\boldsymbol{f}_k\}_{k=1}^K$ when the optimal $\{p_k\}_{k=1}^K$
%are substituted.

%To further simplify our problem, we first optimize the transmit
%power $\{p_k\}_{k=1}^K$, with the transmit precoding vectors
%$\{\boldsymbol{f}_k\}_{k=1}^K$ fixed.

\subsection{Active Beamforming Optimization}
Given $\{\theta_{l,m}\}_{l,m}$, we now discuss how to solve the
active beamforming problem. Note that when the IRS parameters are
fixed, this problem is simplified as a max-min SINR problem for
conventional massive MIMO systems, and its solution has already
been proposed in \cite{CaiQuek11,TanChiang11}. To gain insight
into the problem, let us focus on the transmit power allocation
problem first by assuming the precoding vectors
$\{\boldsymbol{f}_k\}_{k=1}^K$ are fixed. The transmit power
allocation problem can be cast as
\begin{align}
  \max_{\{p_k\}}\min_k\quad &\text{SINR}_k \nonumber \\
  \text{s.t.}\quad & \sum_{k=1}^K p_k \leq P
  \label{Optimization-P2}
\end{align}
By introducing an auxiliary variable $\tau$, the problem
(\ref{Optimization-P2}) can be rewritten as
\begin{align}
  \min_{\{p_k\},\tau}\quad &-\tau \nonumber \\
  \text{s.t.}\quad & \ln\left(\frac{\tau}{\text{SINR}_k}\right)\leq 0, \forall k \nonumber \\
  & \sum_{k=1}^K p_k -P\leq 0
  \label{Optimization-P3}
\end{align}
Using the Karush-Kuhn-Tucker(KKT) conditions and noting that all
inequalities of (\ref{Optimization-P3}) become equalities at the
optimal point, it can be shown \cite{CaiQuek11} that the optimal solution
$\{p_k^0\}_{k=1}^K$ and $\tau^0$ satisfy the following conditions
\begin{align}
  \tau^0 &=\frac{p_k^0\left|\boldsymbol{h}_k^H\boldsymbol{f}_k\right|^2}{\sum_{i\neq k}
  p_i^0\left|\boldsymbol{h}_k^H\boldsymbol{f}_i\right|^2 + \sigma_z^2}, \ \forall k \label{KKT-conditions-1} \\
  \sum_{k=1}^K p_k^0 &= P \label{KKT-conditions-2} \\
  \tau^0&=\frac{q_k^0\left|\boldsymbol{h}_k^H\boldsymbol{f}_k\right|^2}{\sum_{i\neq k} q_i^0
  \left|\boldsymbol{h}_i^H\boldsymbol{f}_k\right|^2 + \sigma_z^2}, \ \forall k \label{KKT-conditions-3} \\
  q_k^0&\triangleq\frac{\sigma_z^2\tau^0\lambda_k^0}{\mu^0p_k^0
  \left|\boldsymbol{h}_k^H\boldsymbol{f}_k\right|^2}, \ \forall k  \label{KKT-conditions-4} \\
  \sum_{k=1}^K \lambda_k^0 &= 1 \label{KKT-conditions-5} \\
  \lambda_k^0 &>0, \ \forall k \label{KKT-conditions-6} \\
  \mu^0 &>0  \label{KKT-conditions-7}
\end{align}
where $\{\lambda_k^0\}$ and $\mu^0$ are optimal Lagrange dual
variables. From (\ref{KKT-conditions-1}),
(\ref{KKT-conditions-2}), and (\ref{KKT-conditions-3}), we can
arrive at
\begin{align}
  \sum_{k=1}^K q_k^0= P
  \label{sum-qk}
\end{align}
The above KKT condition (\ref{KKT-conditions-3}) indicates that
the transmit precoding vector $\boldsymbol{f}_k$ can be found via
\begin{align}
  \max_{\boldsymbol{f}_k}\quad &\frac{q_k^0\left|\boldsymbol{h}_k^H
  \boldsymbol{f}_k\right|^2}{\sum_{i\neq k} q_i^0\left|\boldsymbol{h}_i^H\boldsymbol{f}_k\right|^2 + \sigma_z^2} \nonumber \\
  \text{s.t.}\quad & \left\|\boldsymbol{f}_k\right\|_2=1
  \label{Optimization-P4}
\end{align}
whose optimal solution is given by
\begin{align}
  \boldsymbol{f}_k^0
  =\frac{\left(\sum_{i\neq k}q_i^0\boldsymbol{h}_i \boldsymbol{h}_i^H + \sigma_z^2\boldsymbol{I}\right)^{-1}\boldsymbol{h}_k}
  {\left\|\left(\sum_{i\neq k}q_i^0\boldsymbol{h}_i \boldsymbol{h}_i^H +
  \sigma_z^2\boldsymbol{I}\right)^{-1}\boldsymbol{h}_k\right\|_2}
  \label{optimal-f}
\end{align}
Substituting (\ref{optimal-f}) into (\ref{KKT-conditions-3}), we
have
\begin{align}
  q_k^0 = \frac{\tau^0}
  {\boldsymbol{h}_k^H\left(\sum_{i\neq k}q_i^0\boldsymbol{h}_i \boldsymbol{h}_i^H + \sigma_z^2
  \boldsymbol{I}\right)^{-1}\boldsymbol{h}_k}
  \label{qk}
\end{align}
Based on (\ref{sum-qk}) and (\ref{qk}), $\{q_k^0\}_{k=1}^K$ can be
obtained via a fixed-point iteration method. Specifically, we
first randomly generate $\{q_k^{(1)}\}_{k=1}^K$ as the initial
point. At the $n$th iteration, we calculate
\begin{align}
  q_k'=\frac{1}{\boldsymbol{h}_k^H
\left(\sum_{i\neq k}q_i^{(n)}\boldsymbol{h}_i \boldsymbol{h}_i^H +
\sigma_z^2\boldsymbol{I}\right)^{-1}\boldsymbol{h}_k}
\end{align}
and then update
\begin{align}
  q_k^{(n+1)}=\frac{Pq_k'}{\sum_{k=1}^Kq_k'}
\end{align}
After $\{q_k^0\}_{k=1}^K$ is obtained, $\tau^0$ can then be
calculated from (\ref{sum-qk}) and (\ref{qk}):
\begin{align}
  \tau^0 = \frac{P}{\sum_{k=1}^K\left[\boldsymbol{h}_k^H\left(\sum_{i\neq k}q_i^0\boldsymbol{h}_i
  \boldsymbol{h}_i^H + \sigma_z^2\boldsymbol{I}\right)^{-1}\boldsymbol{h}_k\right]^{-1}}
  \label{tau-opt}
\end{align}
Now, define $\boldsymbol{p}^0\triangleq
[p_1^0\phantom{0}\ldots\phantom{0}p_K^0]^T$,
$\boldsymbol{b}\triangleq\tau^0\sigma_z^2\boldsymbol{1}_K$, and
\begin{align}
  \boldsymbol{A} \triangleq
  \left[
    \begin{array}{cccc}
      \left|\boldsymbol{h}_1^H\boldsymbol{f}_1\right|^2 & -\tau^0
      \left|\boldsymbol{h}_1^H\boldsymbol{f}_2\right|^2 & \cdots & -\tau^0\left|\boldsymbol{h}_1^H\boldsymbol{f}_K\right|^2 \\
      -\tau^0\left|\boldsymbol{h}_2^H\boldsymbol{f}_1\right|^2 &
      \left|\boldsymbol{h}_2^H\boldsymbol{f}_2\right|^2 & \cdots & -
      \tau^0\left|\boldsymbol{h}_2^H\boldsymbol{f}_K\right|^2 \\
      \vdots & \vdots & \ddots & \vdots \\
      -\tau^0\left|\boldsymbol{h}_K^H\boldsymbol{f}_1\right|^2 & -
      \tau^0\left|\boldsymbol{h}_K^H\boldsymbol{f}_2\right|^2 & \cdots & \left|\boldsymbol{h}_K^H\boldsymbol{f}_K\right|^2 \\
    \end{array}
  \right]
  \nonumber
\end{align}
Equation (\ref{KKT-conditions-1}) can be expressed as
$\boldsymbol{A}\boldsymbol{p}^0=\boldsymbol{b}$. Therefore the
optimal transmit power $\boldsymbol{p}^0$ can be obtained as
\begin{align}
  \boldsymbol{p}^0 = \boldsymbol{A}^{-1}\boldsymbol{b}
  \label{optimal-p}
\end{align}

\subsection{Passive Beamforming Optimization}
From (\ref{tau-opt}), the joint beamforming problem
(\ref{Optimization-P1}) can be simplified as a passive beamforming
problem:
\begin{align}
    \max_{\{\theta_{l,m}\}}\quad & \tau^0=\frac{P}{\sum_{k=1}^K
    \left[\boldsymbol{h}_k^H\left(\sum_{i\neq k}q_i^0\boldsymbol{h}_i
  \boldsymbol{h}_i^H + \sigma_z^2\boldsymbol{I}\right)^{-1}\boldsymbol{h}_k\right]^{-1}} \nonumber \\
  \text{s.t.}\quad & \theta_{l,m}\in[0,2\pi), \forall l, \forall m
  \label{Optimization-P4}
\end{align}
Note that in the above optimization, both the composite channel
vectors $\{\boldsymbol{h}_k\}$ and $\{q_k^0\}$ are dependent on
$\{\theta_{l,m}\}$. Our objective is to devise the phase shift
parameters $\{\theta_{l,m}\}$ to maximize the SINR $\tau^0$. Such
a problem, however, is challenging due to the non-convexity of the
objective function. Moreover, $\{q_k^0\}$ cannot be expressed as
an explicit function of $\{\theta_{l,m}\}$, which further
complicates the problem. In the following, we will convert the
above problem into a more amiable form which helps us gain insight
into the problem.

%by exploiting the rank-one structure of the BS-IRS channel,

Substituting (\ref{Gl}) into (\ref{hk}), the composite channel can
be expressed as
\begin{align}
  \boldsymbol{h}_k &= \sum_{l=1}^L \sqrt{NM}\alpha_l^* \boldsymbol{a}_t(\psi_l)
  \boldsymbol{a}_r^H(\varphi_l,\vartheta_l)  \boldsymbol{\Phi}_l^H \boldsymbol{h}_{l,k} \nonumber \\
  &\triangleq \sum_{l=1}^L \sqrt{NM^2}\boldsymbol{a}_t(\psi_l) w_{l,k} \nonumber \\
  &\triangleq \sqrt{NM^2}\boldsymbol{B}\boldsymbol{w}_k
  \label{hk-rankoneGl}
\end{align}
where
\begin{align}
w_{l,k} \triangleq & \alpha_l^*
\boldsymbol{a}_r^H(\varphi_l,\vartheta_l) \boldsymbol{\Phi}_l^H
\boldsymbol{h}_{l,k}/\sqrt{M}  \label{pbg} \\
\boldsymbol{w}_k \triangleq &
[w_{1,k}\phantom{0}\ldots\phantom{0}w_{L,k}]^T
\\
\boldsymbol{B} \triangleq &
[\boldsymbol{a}_t(\psi_1)\phantom{0}\ldots\phantom{0}\boldsymbol{a}_t(\psi_L)]
\end{align}
From (\ref{hk-rankoneGl}), we see that the composite channel from
the BS to the $k$th user, $\boldsymbol{h}_k$, is a linear
combination of the transmit array response vectors
$\{\boldsymbol{a}_t(\psi_l)\}$, with each transmit array response
vector weighted by $w_{l,k}$. The weight $w_{l,k}$ is referred to
as the ``passive beamforming gain'' from the $l$th IRS to the
$k$th user. Substituting (\ref{hk-rankoneGl}) into the objective
function of (\ref{Optimization-P4}), we arrive at
\begin{align}
  \tau^0 &= \frac{P}{\sum_{k=1}^K\left[\boldsymbol{w}_k^H\boldsymbol{B}^H\left(\boldsymbol{B}
  \boldsymbol{R}_{k}\boldsymbol{B}^H +
  \tilde{\sigma}_z^2\boldsymbol{I}\right)^{-1}\boldsymbol{B}\boldsymbol{w}_k\right]^{-1}}
\end{align}
where $\boldsymbol{R}_{k}\triangleq \sum_{i\neq
k}q_i^0\boldsymbol{w}_i\boldsymbol{w}_i^H$, and
$\tilde{\sigma}_z^2\triangleq \sigma_z^2/(NM^2)$.

Since the steering vectors $\{\boldsymbol{a}_t(\psi_l)\}$ are
mutually orthogonal with each other (see A3), there exists a
matrix $\boldsymbol{B}'\in\mathbb{C}^{N\times (N-L)}$ such that
$\boldsymbol{B}_0\triangleq
[\boldsymbol{B}\phantom{0}\boldsymbol{B}']$ is a unitary matrix,
i.e. $\boldsymbol{B}_0^H\boldsymbol{B}_0 =
\boldsymbol{B}_0\boldsymbol{B}_0^H = \boldsymbol{I}$. Using this
fact and the Woodbury matrix identity, we have
\begin{align}
  &\boldsymbol{w}_k^H\boldsymbol{B}^H\left(\boldsymbol{B}\boldsymbol{R}_{k}\boldsymbol{B}^H
  + \tilde{\sigma}_z^2\boldsymbol{I}\right)^{-1}\boldsymbol{B}\boldsymbol{w}_k \nonumber \\
  =\ &\boldsymbol{w}_k^H\boldsymbol{B}^H\left(\boldsymbol{B}_0
  \left[
    \begin{array}{cccc}
      \boldsymbol{R}_{k} & \boldsymbol{0} \\
      \boldsymbol{0} & \boldsymbol{0} \\
    \end{array}
  \right]
  \boldsymbol{B}_0^H
  + \tilde{\sigma}_z^2\boldsymbol{I}\right)^{-1}\boldsymbol{B}\boldsymbol{w}_k \nonumber \\
  =\ &\boldsymbol{w}_k^H\boldsymbol{B}^H\left(\boldsymbol{B}_0
  \left[
    \begin{array}{cccc}
      \boldsymbol{R}_{k} + \tilde{\sigma}_z^2\boldsymbol{I} & \boldsymbol{0} \\
      \boldsymbol{0} & \tilde{\sigma}_z^2\boldsymbol{I} \\
    \end{array}
  \right]
  \boldsymbol{B}_0^H
  \right)^{-1}\boldsymbol{B}\boldsymbol{w}_k \nonumber \\
  =\ &\boldsymbol{w}_k^H\boldsymbol{B}^H\boldsymbol{B}_0
  \left[
    \begin{array}{cccc}
      \left(\boldsymbol{R}_{k} + \tilde{\sigma}_z^2\boldsymbol{I}\right)^{-1} & \boldsymbol{0} \\
      \boldsymbol{0} & \tilde{\sigma}_z^{-2}\boldsymbol{I} \\
    \end{array}
  \right]
  \boldsymbol{B}_0^H\boldsymbol{B}\boldsymbol{w}_k \nonumber \\
  =\ &\boldsymbol{w}_k^H\left(\boldsymbol{R}_{k} + \tilde{\sigma}_z^2\boldsymbol{I}\right)^{-1}\boldsymbol{w}_k \nonumber \\
  \triangleq\ &\boldsymbol{w}_k^H\left(\boldsymbol{W}_{k}\boldsymbol{Q}_{k}\boldsymbol{W}_{k}^H
  + \tilde{\sigma}_z^2\boldsymbol{I}\right)^{-1}\boldsymbol{w}_k \nonumber \\
  =\ &\tilde{\sigma}_z^{-2}\boldsymbol{w}_k^H\left(\boldsymbol{I}-\boldsymbol{W}_{k}
  (\tilde{\sigma}_z^2\boldsymbol{Q}_{k}^{-1}
  +\boldsymbol{W}_{k}^H\boldsymbol{W}_{k})^{-1}\boldsymbol{W}_{k}^H\right)\boldsymbol{w}_k
  \label{Woodbury}
\end{align}
where
\begin{align}
\boldsymbol{W}_{k}\triangleq &
[\boldsymbol{w}_1\phantom{0}\ldots\phantom{0}\boldsymbol{w}_{k-1}\phantom{0}\boldsymbol{w}_{k+1}\phantom{0}
\ldots\phantom{0}\boldsymbol{w}_K] \nonumber\\
\boldsymbol{Q}_{k} \triangleq &
\text{diag}(q_1^0,\ldots,q_{k-1}^0,q_{k+1}^0,\ldots,q_K^0)
\nonumber
\end{align}
Therefore the passive beamforming problem (\ref{Optimization-P4})
becomes
\begin{align}
\max_{\{\theta_{l,m}\}}\quad &   \tau^0 =
\frac{P\tilde{\sigma}_z^{-2}}{\sum_{k=1}^K\left[\boldsymbol{w}_k^H
  (\boldsymbol{I}-\boldsymbol{W}_{k}(\tilde{\sigma}_z^2\boldsymbol{Q}_{k}^{-1}
  +\boldsymbol{W}_{k}^H\boldsymbol{W}_{k})^{-1}\boldsymbol{W}_{k}^H)\boldsymbol{w}_k\right]^{-1}}
  \nonumber\\
 \text{s.t.}\quad & \theta_{l,m}\in[0,2\pi), \quad \forall l, \forall m
 \label{opt5}
\end{align}
Denote
\begin{align}
\boldsymbol{W}\triangleq & [\boldsymbol{w}_1\phantom{0}
\ldots\phantom{0}\boldsymbol{w}_K]
\end{align}
as the passive beamforming matrix with its $(l,k)$th entry given
by $w_{l,k}$. Clearly, to maximize the SINR $\tau_0$, on one hand,
we wish to make the passive beamforming gains $\{w_{l,k}\}$ as
large as possible; on the other hand, we wish
$\{\boldsymbol{w}_k\}$ to be orthogonal to each other such that
the cross-interference term
$\boldsymbol{w}_k^H\boldsymbol{W}_{k}(\tilde{\sigma}_z^2\boldsymbol{Q}_{k}^{-1}
  +\boldsymbol{W}_{k}^H\boldsymbol{W}_{k})^{-1}\boldsymbol{W}_{k}^H\boldsymbol{w}_k$
is minimized. A key insight here is that when the phase shift
parameters of the $l$th IRS are optimally tuned to serve the $k$th
user, i.e. $|w_{l,k}|$ is maximized, due to the asymptotic
orthogonality among channel vectors associated with different
users, this IRS will be like non-existent and interference-free to
other users, i.e. $w_{l,k'}\approx 0$ for $k'\neq k$. In this
case, each row of the passive beamforming matrix $\boldsymbol{W}$
contains only a single nonzero element, and as a result, its
columns $\{\boldsymbol{w}_k\}$ are mutually orthogonal. Such a
property, termed as automatic interference cancelation (AIC), will
be rigorously proved in Section \ref{sec:AIC}. This property is
important as it can help avoid complicated inter-IRS interference
management.

Based on the AIC property, a simple yet effective solution is to
divide $L$ IRSs into $K$ disjoint groups, with parameters of each
group of IRSs optimally tuned to serve one particular user. This
problem is referred to as a user association problem. This
solution is effective because, on one hand, it can achieve a
reasonably large $\|\boldsymbol{w}_k\|_2^2$ for each $k$; on the
other hand, the cross-interference term disappears due to the
orthogonality among $\{\boldsymbol{w}_k\}$. To formulate this user
association problem, define $w_{l,k}^{\star}$ as the maximum
achievable passive beamforming gain. From (\ref{pbg}), it can be
easily verified that $|w_{l,k}^{\star}|$ is given by
\begin{align}
|w_{l,k}^{\star}|=\left|\frac{\alpha_l^*}{M}\right|\sum_{m=1}^{M}|\boldsymbol{h}_{l,k}(m)|
\end{align}
which is attained when parameters associated with the $l$th IRS
are set as
\begin{align}
  \theta_{l,m} &=
  \arg(\boldsymbol{h}_{l,k}(m))-\arg(\boldsymbol{a}_{r}(m)) \quad
\forall m \label{theta-opt}
\end{align}
where $\arg(x)$ represents the argument of the complex number $x$,
$\boldsymbol{h}_{l,k}(m)$ and $\boldsymbol{a}_{r}(m)$ denote the
$m$th entry of $\boldsymbol{h}_{l,k}$ and
$\boldsymbol{a}_{r}(\varphi_l,\vartheta_l)$, respectively. Let
$\boldsymbol{W}^{\star}\triangleq\{w_{l,k}^{\star}\}$. The
IRS-user association problem can be formulated as
\begin{align}
\min_{\boldsymbol{W}}\quad &
\frac{1}{\tau^{0}}=\frac{\tilde{\sigma}_z^2}{P}\sum_{k=1}^K\frac{1}{\|\boldsymbol{w}_k\|_2^2}
  \nonumber\\
 \text{s.t.}\quad & \|\boldsymbol{W}[l,:]\|_0=1, \quad \forall l, \nonumber \\
 & w_{l,k} \in\left\{0,w_{l,k}^{\star}\right\}, \quad \forall l, \forall k
 \label{opt6}
\end{align}
where entries of $\boldsymbol{W}$ are chosen either to be
$w_{l,k}^{\star}$ or 0, $\boldsymbol{W}[l,:]$ denotes the $l$th
row of $\boldsymbol{W}$, and the constraint
$\|\boldsymbol{W}[l,:]\|_0=1$ is due to the AIC property, i.e.
when the phase shift parameters of the $l$th IRS are optimally
tuned to serve a certain user, this IRS will become
interference-free to other users. Altogether, $\boldsymbol{W}$ is
constructed by keeping only one element for each row of
$\boldsymbol{W}^{\star}$ and setting the rest elements of
$\boldsymbol{W}^{\star}$ equal to zero. Since we can choose any
element from each row of $\boldsymbol{W}^{\star}$, the number of
feasible solutions is up to $K^{L}$, from which we need to choose
the one that maximizes the objective function in (\ref{opt6}).

The simplest method to solve (\ref{opt6}) is to exhaustively
search all feasible solutions. When $K$ and $L$ are small, say
$L\leq 6$, the computational complexity of this exhaustive search
scheme is still practical. For the scenario where a large number
of IRSs are deployed, we develop a greedy algorithm to search for
the best IRS-user association. Specifically, we first choose the
largest passive beamforming gain from $\boldsymbol{W}^{\star}$,
say $w_{l_1,k_1}^{\star}$ is the largest (in terms of magnitude)
in $\boldsymbol{W}^{\star}$. Based on this result, the $l_1$th IRS
is allocated to the $k_1$th user. Next, we nullify the $l_1$th IRS
and the $k_1$th user by setting entries on the $l_1$th row and the
$k_1$th column of $\boldsymbol{W}^{\star}$ equal to zero. Thus we
obtain an updated passive beamforming gain matrix, denoted as
$\boldsymbol{W}_{1}^{\star}$. Then we choose the largest element
(in terms of magnitude) in $\boldsymbol{W}_{1}^{\star}$, say
$w_{l_2,k_2}^{\star}$, and let the $l_2$th IRS assigned to the
$k_2$th user. Again, the passive beamforming gain matrix is
updated by setting entries of the $l_2$th row and the $k_2$th
column of $\boldsymbol{W}_{1}^{\star}$ equal to zero. This
procedure is repeated until each user is served by an IRS. If the
number of IRSs is larger than the number of users, then we need to
assign extra IRSs to users. We first set the
$\{l_1,\ldots,l_k\}$th rows of $\boldsymbol{W}^{\star}$ equal to
zero and obtain a new matrix $\boldsymbol{W}_{k+1}^{\star}$. Then,
we choose the largest entry (in terms of magnitude) in
$\boldsymbol{W}_{k+1}^{\star}$, say
$w_{l_{k+1},\bar{k}_{1}}^{\star}$, and let the $l_{k+1}$th IRS
assigned to the $\bar{k}_{1}$th user. Next, the passive
beamforming gain matrix is updated by setting the $l_{k+1}$th row
of $\boldsymbol{W}_{k+1}^{\star}$ equal to zero. This procedure is
repeated until all IRSs are assigned. Although this greedy
algorithm is not guaranteed to yield the optimal solution, it has
a very low computational complexity.

%find the IRS-user pair $\{l_1,k_1\}$ which achieves the largest
%passive beamforming gain $w_{l_1,k_1}^{\star}$ from all possible
%pairs $\{l,k\}$, and then tune the parameters of the $l_1$th IRS
%to serve the $k_1$th user. Since the $l_1$th IRS has already been
%tuned to serve the $k_1$th user, we need to exclude the $l_1$th
%IRS when finding the best IRS-user pair in the following steps.
%Next, for the rest of $K-1$ users, we again find the IRS-user pair
%$\{l_2,k_2\}$ with the largest $w_{l_2,k_2}^{\star}$, and then
%tune the parameters of the $l_2$th IRS to serve the $k_2$th user.
%This procedure repeats until each user is served by an IRS.

\begin{algorithm}[!h]
\caption{Greedy Search Algorithm}
\begin{algorithmic}
\STATE { Given the channel $\boldsymbol{G}_l$ and
$\boldsymbol{h}_{l,k}$, $\forall l,k$. \STATE Define
$\mathcal{L}=\{1,\ldots,L\}$, and $\mathcal{K}=\{1,\ldots,K\}$.
\STATE Initialize
$\mathcal{\tilde{L}}=\mathcal{\tilde{K}}=\varnothing$.
\FOR{$i=1,\ldots,K$} \STATE Find
$\{l_i,k_i\}=\mathop{\arg\max}_{l\in
\mathcal{L}-\mathcal{\tilde{L}}, k\in
\mathcal{K}-\mathcal{\tilde{K}}}w_{l,k}^{\star}$, and then design
$\theta_{l_i,m}$ via (\ref{theta-opt}) such that
$|w_{l_i,k_i}|=w_{l_i,k_i}^{\star}$. \STATE Let
$\mathcal{\tilde{L}} = \mathcal{\tilde{L}} \cap \{l_j\}$, and
$\mathcal{\tilde{K}} = \mathcal{\tilde{K}} \cap \{k_j\}$.
 \ENDFOR \IF
{$L>K$} \FOR{$i=K+1,\ldots,L$} \STATE Find
$\left\{l_i,k_i\right\}=\mathop{\arg\max}_{l\in
\mathcal{L}-\mathcal{\tilde{L}},k\in
\mathcal{K},}w_{l,k}^{\star}$, and then design $\theta_{l_i,m}$
via (\ref{theta-opt}) such that
$|w_{l_i,k_i}|=w_{l_i,k_i}^{\star}$. \STATE Let
$\mathcal{\tilde{L}} = \mathcal{\tilde{L}} \cap \{l_i\}$. \ENDFOR
\ENDIF }
\end{algorithmic}
\label{algorithm1}
\end{algorithm}

\section{Automatic Interference Cancelation} \label{sec:AIC}
In the previous section, we propose an effective solution that
converts the passive beamforming problem into a user association
problem, based on the property that when the $l$th IRS is
optimally tuned to serve the $k$th user, this IRS will become
non-existent (i.e. interference-free) to other users. Such a
property is referred to as the AIC property. In this section, we
consider two typical IRS-user channel models, namely, an
LOS-dominated channel model and a Rayleigh fading channel model.
We will show that the AIC property holds asymptotically for both
scenarios when $M\rightarrow\infty$.

\subsection{LOS-Dominated IRS-User Channel}
Suppose the channel between the $l$th IRS and the $k$th user
$\boldsymbol{h}_{l,k}$ ($\forall l,k$) is dominated by the LOS
component, which is usually the case for mmWave communications \cite{HurKim13}.
In this case, $\boldsymbol{h}_{l,k}$ can be written as
\begin{align}
  \boldsymbol{h}_{l,k} = \beta_{l,k} \sqrt{M}\boldsymbol{a}_{r}(\phi_{l,k},\omega_{l,k})
  \label{LOS-channel}
\end{align}
where $\beta_{l,k}$ is the complex gain between the $l$th IRS and
the $k$th user, $\phi_{l,k}\in [0, 2\pi]$ ($\omega_{l,k}\in [0,
2\pi]$) is the associated azimuth (elevation) AoD, and
$\boldsymbol{a}_{r}\in\mathbb{C}^{M}$ is the array response vector
associated with the IRS. Assume a uniform planar array (UPA) is
used at each IRS, the steering vector can be expressed as
\begin{align}
  &\boldsymbol{a}_{r}(\phi_{l,k},\omega_{l,k}) \nonumber \\
  =&\frac{1}{\sqrt{M}}[1\phantom{0}\ldots\phantom{0}e^{j\frac{2\pi}{\lambda}d\left((m_1-1)
  \cos(\omega_{l,k})\sin(\phi_{l,k})+(m_2-1)\sin(\omega_{l,k})\right)}\phantom{0}\nonumber \\
  &\ldots\phantom{0}e^{j\frac{2\pi}{\lambda}d\left((M_y-1)\cos(\omega_{l,k})
  \sin(\phi_{l,k})+(M_z-1)\sin(\omega_{l,k})\right)}]^T \nonumber
\end{align}
where $M_y$ ($M_z$) denotes the number of elements along the
horizontal (vertical) axis, $M=M_y M_z$, and $(m_1, m_2)$ is the
coordinate of the reflecting element. From (\ref{pbg}), we know
that $w_{l,k}$ is given by
\begin{align}
   & w_{l,k}  \nonumber \\
  =& \frac{\alpha_l^*}{\sqrt{M}}\boldsymbol{a}_r^H(\varphi_l,\vartheta_l)
  \boldsymbol{\Phi}_l^H \boldsymbol{h}_{l,k} \nonumber \\
  =& \alpha_l^*\beta_{l,k}\boldsymbol{a}_r^H(\varphi_l,\vartheta_l)
  \boldsymbol{\Phi}_l^H \boldsymbol{a}_{r}(\phi_{l,k},\omega_{l,k}) \nonumber \\
  =& \frac{\alpha_l^*\beta_{l,k}}{M}\sum_{m_1=1}^{M_y}
  \sum_{m_2=1}^{M_z}\exp\{-j\theta_{l,m_1,m_2}+ \nonumber \\
  &j\frac{2\pi d}{\lambda}((m_1-1)\left(\cos(\omega_{l,k})
  \sin(\phi_{l,k})-\cos(\vartheta_{l})\sin(\varphi_{l})\right) \nonumber \\
  &+(m_2-1)\left(\sin(\omega_{l,k})-\sin(\vartheta_{l})\right))\} \nonumber \\
  \triangleq& \frac{\alpha_l^*\beta_{l,k}}{M}\sum_{m_1=1}^{M_y}
  \sum_{m_2=1}^{M_z}\exp\{-j\theta_{l,m_1,m_2}+j\mu_{m_1,m_2,l,k}\}
  \label{w-kl}
\end{align}
where $\theta_{l,m_1,m_2}$ denotes the phase shift parameter of
the $l$th IRS's reflecting element at the coordinate $(m_1, m_2)$,
and
\begin{align}
\mu_{m_1,m_2,l,k}\triangleq&\frac{2\pi
d}{\lambda}\big((m_1-1)(\cos(\omega_{l,k})\sin(\phi_{l,k})-\cos(\vartheta_{l})\sin(\varphi_{l}))
\nonumber\\
&+(m_2-1)(\sin(\omega_{l,k})-\sin(\vartheta_{l}))\big)
\end{align}
It is clear that $|w_{l,k}|$ is maximized when
\begin{align}
  \theta_{l,m_1,m_2} = \mu_{m_1,m_2,l,k}, \quad \forall m_1, \forall m_2
  \label{LoS-theta}
\end{align}
When the parameters of the $l$th IRS are optimally tuned to serve
the $k$th user, we now show that this IRS becomes
interference-free to other users in an asymptotic sense, i.e.
$|w_{l,k'}|\rightarrow 0$ as $M\rightarrow\infty$ for any $k'\neq
k$. We have $|w_{l,k'}|(\forall k'\neq k)$
\begin{align}
   &|w_{l,k'}| \nonumber \\
  =& \left|\frac{\alpha_l^*}{\sqrt{M}}\boldsymbol{a}_r^H(\varphi_l,\vartheta_l)
  \boldsymbol{\Phi}_l^H \boldsymbol{h}_{l,k'}\right| \nonumber \\
  =& \left|\frac{\alpha_l^*\beta_{l,k'}}{M}\sum_{m_1=1}^{M_y}
  \sum_{m_2=1}^{M_z}\exp\{-j\mu_{m_1,m_2,l,k}+j\mu_{m_1,m_2,l,k'}\}\right| \nonumber \\
  =& \left|\alpha_l^*\beta_{l,k'}\right|\frac{\text{sinc}
  \left(\frac{\pi d}{\lambda}M_y\delta_{l,k,k'}\right)}{\text{sinc}
  \left(\frac{\pi d}{\lambda}\delta_{l,k,k'}\right)}\frac{\text{sinc}
  \left(\frac{\pi d}{\lambda}M_z\gamma_{l,k,k'}\right)}{\text{sinc}\left(\frac{\pi d}{\lambda}\gamma_{l,k,k'}\right)}
\end{align}
where
\begin{align}
  \text{sinc}(x)&\triangleq \frac{\sin(x)}{x} \nonumber \\
  \delta_{l,k,k'}&\triangleq \cos(\omega_{l,k'})\sin(\phi_{l,k'})-\cos(\omega_{l,k})\sin(\phi_{l,k}) \nonumber \\
  \gamma_{l,k,k'}&\triangleq \sin(\omega_{l,k'})-\sin(\omega_{l,k})
\end{align}
Suppose the locations of different users are well separated such
that $\delta_{l,k,k'}\neq 0$ and $\gamma_{l,k,k'}\neq 0$, $\forall
k'\neq k$. Then we have
\begin{align}
  \lim_{M\rightarrow \infty} |w_{l,k'}| = 0, \quad \forall k'\neq k
\end{align}
Hence the AIC property holds in an asymptotic sense as
$M\rightarrow\infty$.

%and the radio signal is scattered by a number of objects before it
%arrives at the user

%Suppose the coefficients of channels $\{\boldsymbol{h}_{l,k}\}$
%for different IRSs or users are independent.

%In the following, we will show that if the parameters associated
%with the $l$th IRS, i.e. $\{\theta_{l,m}\}$, are optimally devised
%to maximize $|w_{l,k}|$, then $|w_{l,k'}| (\forall k'\neq k)$
%approaches zero asymptotically as $M$ goes to infinity.

\subsection{Rayleigh Channel}
For the scenario where there is no dominant propagation along an
LOS between the IRS and the user, the channel can be modeled as an
independent and identically distributed (i.i.d.) Rayleigh fading
channel \cite{WuZhang19b}, i.e.
\begin{align}
  \boldsymbol{h}_{l,k} \sim \mathcal{CN}(\boldsymbol{0},\zeta_{l,k}\boldsymbol{I}) %\tilde{h}_{k,l}(i)
  \label{Rayleigh-channel}
\end{align}
where $\zeta_{l,k}$ is a factor that depends on the distance
between the $l$th IRS and the $k$th user. According to
(\ref{theta-opt}), the passive beamforming gain $|w_{l,k}|$ is
maximized when
\begin{align}
  \theta_{l,m} &=
  \arg(\boldsymbol{h}_{l,k}(m))-\arg(\boldsymbol{a}_{r}(m)) \quad
\forall m
\end{align}
where $\boldsymbol{h}_{l,k}(m)$ and $\boldsymbol{a}_{r}(m)$ denote
the $m$th entry of $\boldsymbol{h}_{l,k}$ and
$\boldsymbol{a}_{r}(\varphi_l,\vartheta_l)$, respectively. The
$|w_{l,k'}|(\forall k'\neq k)$ can then be calculated as
\begin{align}
  |w_{l,k'}| &= \left|\frac{\alpha_l^*}{\sqrt{M}}\boldsymbol{a}_r^H(\varphi_l,\vartheta_l)
  \boldsymbol{\Phi}_l^H \boldsymbol{h}_{l,k'}\right| \nonumber \\
  &= \left|\frac{\alpha_l^*}{M}\sum_{m=1}^{M}
  e^{-j \arg(\boldsymbol{h}_{l,k}(m))}\boldsymbol{h}_{l,k'}(m)\right| \nonumber \\
  &\triangleq \left|\frac{\alpha_l^*}{M}\sum_{m=1}^{M}X_{l,k,k',m}\right|
\end{align}
where $X_{l,k,k',m}\triangleq e^{-j
\arg(\boldsymbol{h}_{l,k}(m))}\boldsymbol{h}_{l,k'}(m)$. Since
$\{X_{l,k,k',1},\ldots,X_{l,k,k',M}\}$ are i.i.d. with
$E[X_{l,k,k',m}]=0$, according to Khinchin's law of large numbers,
we have
\begin{align}
  \lim_{M\rightarrow \infty} P\left\{\left|\frac{1}{M}\sum_{m=1}^{M}X_{l,k,k',m}\right|<\epsilon\right\}=1
\end{align}
for any $\epsilon>0$, i.e.
\begin{align}
  |w_{l,k'}|\rightarrow 0, \quad \forall k'\neq k
\end{align}
when $M\rightarrow \infty$. Therefore, the AIC property holds
asymptotically for the Rayleigh channel.

%when the optimal beamforming vectors $\{\boldsymbol{f}_k^0\}$, the
%optimal power allocation $\{p_k^0\}$, and the proposed IRS
%parameters design are employed on all IRSs and users

\section{Performance Analysis} \label{sec:theoretical-analysis}
In this section, we provide a theoretical analysis of our proposed
method. Specifically, given the global channel state information
$\{\boldsymbol{G}_l\}$ and $\{\boldsymbol{h}_{l,k}\}$, we analyze
the max-min SINR attained by our proposed solution. According to
(\ref{opt5}), the max-min SINR is given by
\begin{align}
  \text{SINR}_k = \tau_0
  =&  \frac{P NM^2\sigma_z^{-2}}{\sum_{k=1}^K\frac{1}{\boldsymbol{w}_k^H
  \left(\boldsymbol{I}-\boldsymbol{W}_{k}\left(\tilde{\sigma}_z^2\boldsymbol{Q}_{k}^{-1}
  +\boldsymbol{W}_{k}^H\boldsymbol{W}_{k}\right)^{-1}\boldsymbol{W}_{k}^H\right)\boldsymbol{w}_k}}
\end{align}
In the previous section, we have shown that the AIC property holds
asymptotically for both LOS-dominated and Rayleigh fading
channels. In other words, when the parameters of the $l$th IRS are
optimally tuned to serve the $k$th user, this IRS is
interference-free to other users. As a result, vectors
$\{\boldsymbol{w}_k\}_{k=1}^K$ are orthogonal to each other, and
we have
\begin{align}
  \tau_0 \approx& \frac{P NM^2\sigma_z^{-2}}{\sum_{k=1}^K\left(\|\boldsymbol{w}_k\|_2^2\right)^{-1}} \nonumber \\
  =&\frac{PNM^2\sigma_z^{-2}}{\sum_{k=1}^K\left(\sum_{i=1}^{n_k}|w^{\star}_{l_i^{(k)},k}|^2 \right)^{-1}} \nonumber \\
  \triangleq& \frac{PNM^2\sigma_z^{-2}}{\sum_{k=1}^K\left(\sum_{i=1}^{n_k}
  \left|\alpha_{l_i^{(k)}}x_{l_i^{(k)},k}\right|^2 \right)^{-1}}
\end{align}
where $\{l_1^{(k)},\ldots,l_{n_k}^{(k)}\}$ is the set of indices
of IRSs that serve the $k$th user, $n_k$ is the number of IRSs
that serve the $k$th user with $\sum_{k=1}^Kn_k=L$, and
\begin{align}
  x_{l,k} \triangleq \frac{1}{M}\sum_{m=1}^{M}|\boldsymbol{h}_{l,k}(m)|
\end{align}
is a constant that depends on the realization of the channel
$\boldsymbol{h}_{l,k}$. Suppose the channel $\boldsymbol{h}_{l,k}$
is a LOS-dominated channel given by (\ref{LOS-channel}). Then
$\tau_0$ can be further calculated as
\begin{align}
  \tau_0  \approx \frac{PNM^2\sigma_z^{-2}}{\sum_{k=1}^K\left(\sum_{i=1}^{n_k}
  \left|\alpha_{l_i^{(k)}}\beta_{l_i^{(k)},k}\right|^2 \right)^{-1}}
  \label{SINR-theo}
\end{align}
From (\ref{SINR-theo}), we can see that the SINR is in the order
of $\mathcal{O}(NM^2)$, which scales quadratically with the number
of reflecting elements $M$. Such a ``squared improvement'' has
also been reported in previous IRS-assisted works, e.g.
\cite{WuZhang18,WangFang19a}. Nevertheless, to our best knowledge,
our work seems to be the first to show that the squared
improvement also holds valid for multi-user systems. Also,
(\ref{SINR-theo}) suggests that, even with a moderate number of
active antennas $N$ at the BS, increasing the number of passive
elements $M$ can help achieve a massive MIMO like gain, thereby
significantly reducing the energy consumption at the BS.

%This result implies that scaling up the number of reflecting
%elements is a promising way to compensate for the significant path
%loss in mmWave wireless communications.

%we can reduce the number of active antennas $N$ at the cost of
%increasing the number of passive elements at IRSs, can be reduced

%For the special case $K=1$, $\tau_0$ is simplified as

%This result suggests that when the direct paths between the
%transmitter and the users are blocked or suffer from high path
%loss, the use of IRSs-assisted system has the potential to provide
%considerable passive beamforming gain for users in massive MIMO
%systems.

\begin{figure}[!t]
\centering
\includegraphics[width=3.5in]{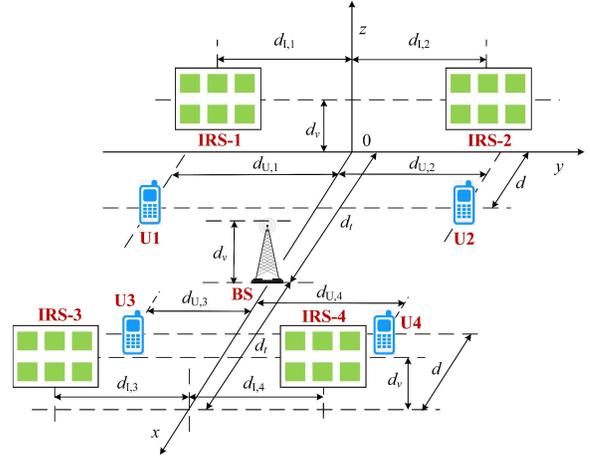}
\caption{Simulation setup 1.} \label{fig-setup-1}
\end{figure}

\section{Simulation Results} \label{sec:simulation-results}
We now provide simulation results to illustrate the performance of
our proposed joint active and passive beamforming solution. In our
simulations, the BS employs a ULA with $N$ antennas, and each IRS
consists of a uniform planar array (UPA) with $M = M_yM_z$
reflecting elements, where $M_y$ and $M_z$ denote the number of
elements along the horizontal axis and vertical axis,
respectively. Throughout our simulations, we fix $M_y=20$, and
increase $M_z$ to obtain different values of $M$. The channel from
the BS to the $l$th IRS $\boldsymbol{G}_l$ is characterized by the
rank-one geometric channel model (\ref{Gl}), in which the complex
gain $\alpha_l$ is generated according to a complex Gaussian
distribution
\begin{align}
  \alpha_l \sim \mathcal{CN}(0,\kappa_l)
\end{align}
where $\kappa_l$ is characterized by a distance-dependent path
loss model given by \cite{ErcegGreenstein99}
\begin{align}
  \kappa_l = C_0 \left(\frac{d_l}{D_0}\right)^{-a_{\text{LOS}}}
\end{align}
$d_l$ is the distance between the BS and the $l$th IRS, $C_0$ is
the path loss at the reference distance $D_0=1$ meter, and
$a_{\text{LOS}}$ denotes the path loss exponent of the
LOS-dominant channel. The channel between the $l$th IRS and the
$k$th user $\boldsymbol{h}_{l,k}$ ($\forall l,k$) is given by
(\ref{LOS-channel}), in which the complex gain $\beta_{l,k}$ is
generated according to a complex Gaussian distribution
\begin{align}
  \beta_{l,k} \sim \mathcal{CN}(0,\rho_{l,k})
\end{align}
where $\beta_{l,k}$ is given by
\begin{align}
  \rho_{l,k} = C_0 \left(\frac{d_{l,k}}{D_0}\right)^{-a_{\text{LOS}}}
\end{align}
$d_{l,k}$ is the distance between the $l$th IRS and the $k$th
user. Some related parameters are set as follows: $P = -10$dBm,
$\sigma_z^2 = -80$dBm, $C_0 = -30$dB, and $a_{\text{LOS}} = 2$.

%The antenna gain $\lambda_t$ and $\lambda_r$ at the BS and each
%user are set to $0$dBi, while the antenna gain $\lambda_I$ at each
%IRS is set to $5$dBi.

We first examine the validity of the AIC property. Consider a
three-dimensional setup, where four IRSs are used to serve four
users, see Fig. \ref{fig-setup-1}. The BS is located on the
$x$-axis with its coordinate given by $(d_t,0,d_v)$, where we set
$d_t=30$m and $d_v=0.3$m. The four IRSs, named as IRS-1, IRS-2,
IRS-3 and IRS-4, are located at $(0,-d_{\text{I},1},d_v)$,
$(0,d_{\text{I},2},d_v)$, $(2d_t,-d_{\text{I},3},d_v)$, and
$(2d_t,d_{\text{I},4},d_v)$, respectively, and we set
$d_{\text{I},1}=d_{\text{I},2}=5$m,
$d_{\text{I},3}=d_{\text{I},4}=3$m. The coordinates of the four
users, namely U1, U2, U3, and U4, are set to
$(d,-d_{\text{U},1},0)$, $(d,d_{\text{U},2},0)$,
$(2d_t-d,-d_{\text{U},3},0)$, and $(2d_t-d,d_{\text{U},4},0)$,
respectively, where $d_{\text{U},1}$, $d_{\text{U},2}$,
$d_{\text{U},3}$, and $d_{\text{U},4}$ are uniformly generated
from $[0,10]$m. To validate the AIC property, we compare our
proposed joint active and passive beamforming solution with the
theoretical result (\ref{SINR-theo}) obtained by neglecting the
cross-interference terms.

\begin{figure}[!t]
\centering
\includegraphics[width=3.5in]{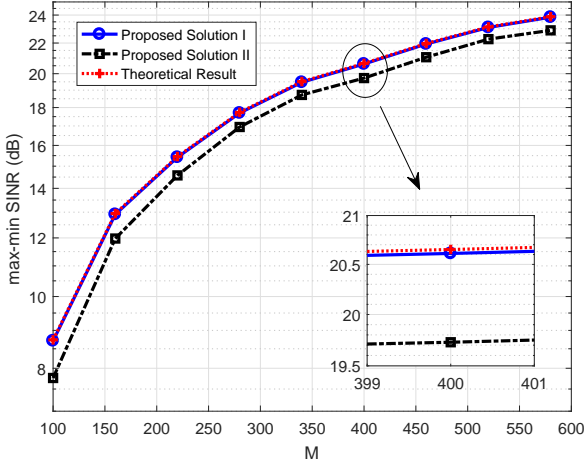}
\caption{Max-min SINRs of respective solutions vs. $M$.}
\label{fig-SINRvsM}
\end{figure}

\begin{figure}[!t]
\centering
\includegraphics[width=3.5in]{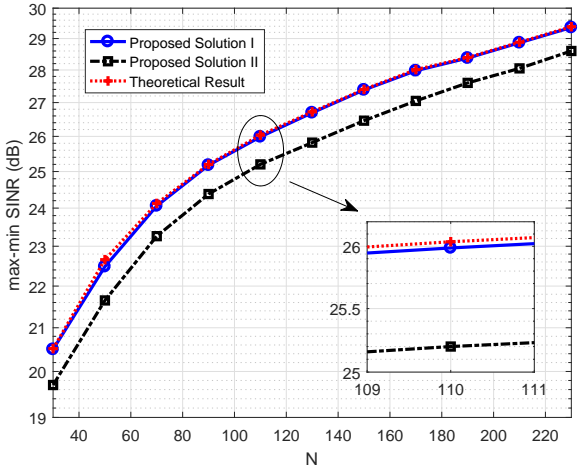}
\caption{Max-min SINRs of respective solutions vs. $N$.}
\label{fig-SINRvsN}
\end{figure}

%For both our solution and the theoretical result
%(\ref{SINR-theo}), the exhaustive search scheme is employed to
%solve the user association problem ().

%We also consider an ideal scenario where the BS serves each user
%separately, as if users are served via orthogonal frequency or
%time resources. The problem therefore is simplified as a set of
%independent joint beamforming problem for MISO systems whose
%optimal solution can be found in [].

%For a fair comparison, the transmit power and the IRSs assigned
%for each user are the same as our proposed solution.

%It is clear that the $k$th user's SINR for this ideal scenario is
%given by
%\begin{align}
%  \max_{\{\boldsymbol{f}_k\}} \frac{p_k^0\left|\boldsymbol{h}_k^H\boldsymbol{f}_k\right|^2}{\sigma_z^2}
%\end{align}
%where $p_k^0$ and $\boldsymbol{h}_k$ are the same as our proposed
%solution

%

Fig. \ref{fig-SINRvsM} plots the max-min SINR of our proposed
solutions as a function of the number of reflecting elements $M$,
where we set $d=5$m and $N=32$. For the theoretical result
(\ref{SINR-theo}) and the proposed solution I, an exhaustive
search scheme is employed to solve the user association problem
(\ref{opt6}), while the proposed solution II uses the greedy
algorithm to solve (\ref{opt6}). From Fig. \ref{fig-SINRvsM}, we
see that doubling the number of reflecting elements achieves a
gain of about $6$dB, which corroborates our theoretical analysis
that the max-min SINR increases quadratically with the number of
reflecting elements. In addition, from Fig. \ref{fig-SINRvsM} we
see that our proposed solution I attains performance close to the
theoretical result (\ref{SINR-theo}). This result indicates that
the AIC property holds well even for a moderate number of
reflecting elements. In Fig. \ref{fig-SINRvsN}, we plot the
max-min SINR of our proposed solutions as a function of $N$, where
we set $d=5$m and $M=400$. From Fig. \ref{fig-SINRvsN}, we see
that doubling the number of antennas at the BS leads to about
$3$dB gain, which corroborates our theoretical result that the
max-min SINR increases linearly with the number of antennas at the
BS. Also, our proposed solution I coincides well with the
theoretical result (\ref{SINR-theo}) for different values of $N$.
This is because the AIC property holds irrespective of the choice
of $N$.

\begin{figure}[!t]
\centering
\includegraphics[width=3.5in]{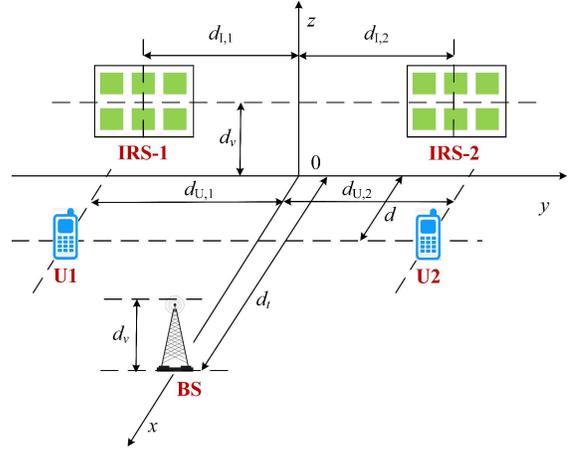}
\caption{Simulation setup 2.} \label{fig-setup-2}
\end{figure}

%where $\{q_k\}$ can be calculated by the iterative method given by
%(\ref{qk-iter1}) and (\ref{qk-iter2}).

\begin{figure}[!t]
\centering
\includegraphics[width=3.5in]{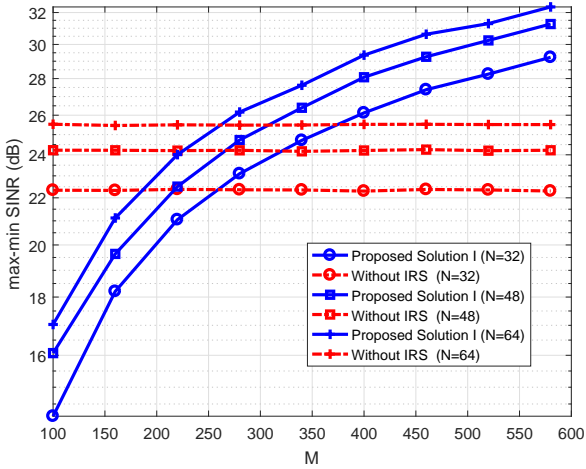}
\caption{Max-min SINRs of IRS-assisted and conventional systems
vs. $M$.} \label{fig-SINRvsM-2I2U}
\end{figure}

\begin{figure}[!t]
\centering
\includegraphics[width=3.5in]{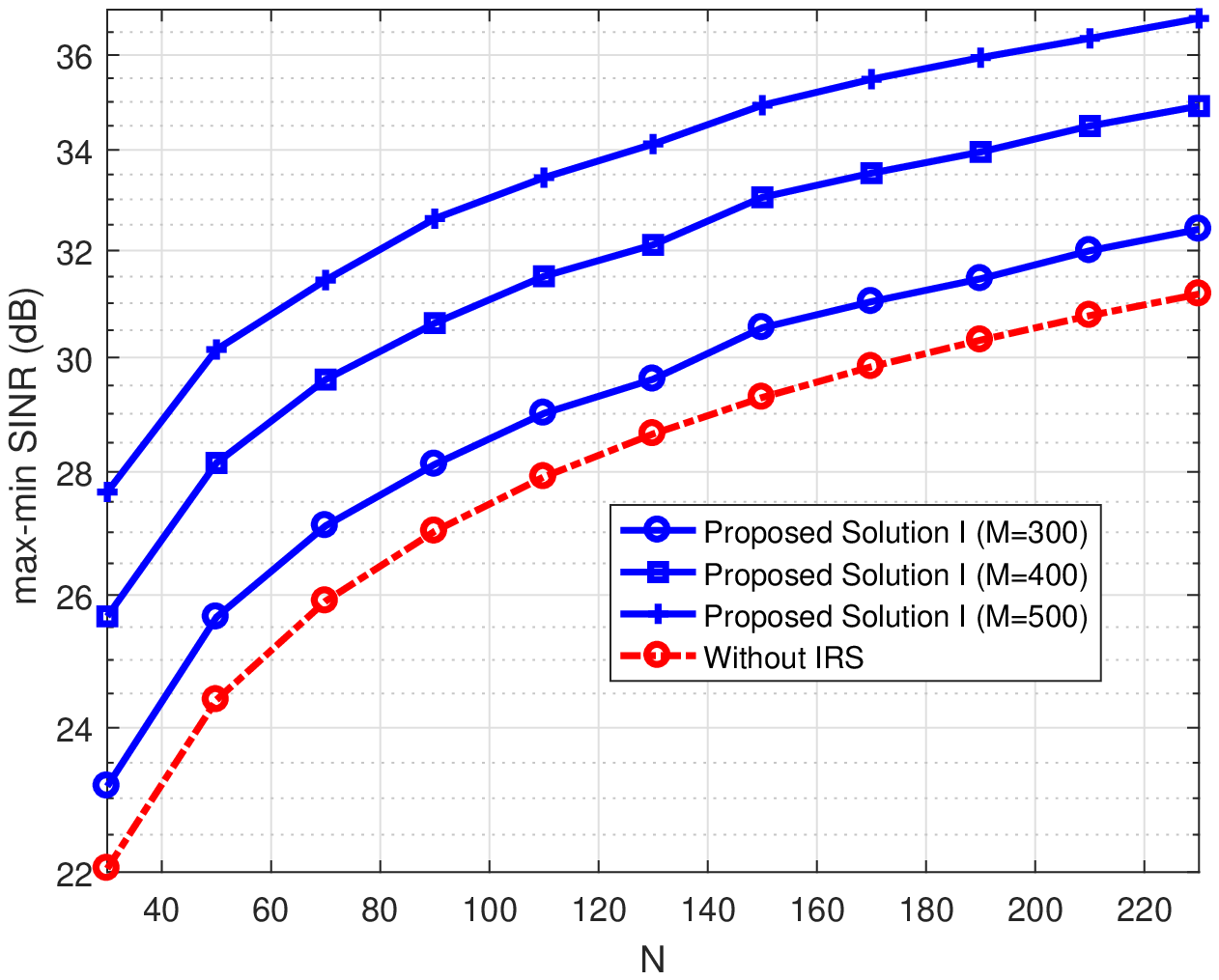}
\caption{Max-min SINRs of IRS-assisted and conventional systems
vs. $N$.} \label{fig-SINRvsN-2I2U}
\end{figure}

\begin{figure}[!t]
\centering
\includegraphics[width=3.5in]{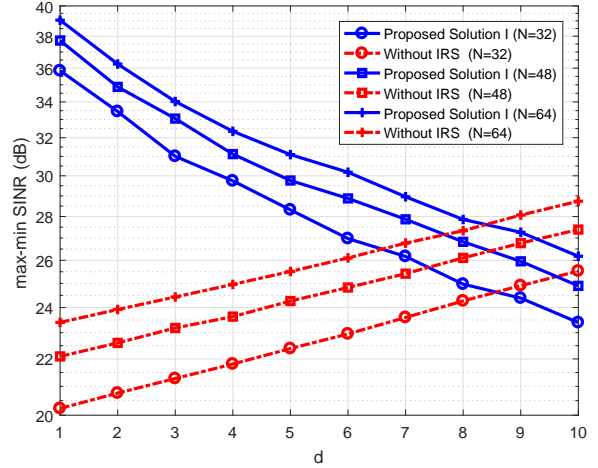}
\caption{Max-min SINRs of IRS-assisted and conventional systems
vs. $d$.} \label{fig-SINRvsd-2I2U}
\end{figure}

Next, we compare our proposed solution with a conventional massive
MIMO system without deploying IRSs. For the conventional massive
MIMO system, the channel between the BS and the $k$th user is
characterized by a geometric channel
\begin{align}
  \tilde{\boldsymbol{h}}_k = \sqrt{N}\sum_{l=1}^{\tilde{L}}
  \tilde{\alpha}_l\boldsymbol{a}_t^H(\tilde{\psi}_l)
  \label{channel-withoutIRS}
\end{align}
where $\tilde{\alpha}_l$ represents the complex gain of the $l$th
path, $\tilde{L}=100$ is the number of paths, and
$\tilde{\psi}_l\in[-\pi/2,\pi/2]$ is the azimuth AoD of the $l$th
path. The complex gain $\tilde{\alpha}_l$ is generated from the
following complex Gaussian distribution
\begin{align}
  \tilde{\alpha}_l \sim \mathcal{CN}(0,\tilde{\kappa}_k)
\end{align}
where $\tilde{\kappa}_k$ is given by
\begin{align}
  \tilde{\kappa}_k = C_0 \left(\frac{\tilde{d}_k}{D_0}\right)^{-a_{\text{NLOS}}}
\end{align}
in which the path loss exponent $a_{\text{NLOS}}$ is set to $3.5$,
and $\tilde{d}_k$ denotes the distance between the BS and the
$k$th user. Note that for the conventional massive MIMO system,
the max-min SINR can be obtained by substituting
(\ref{channel-withoutIRS}) into (\ref{tau-opt}). The simulation
setup is depicted in Fig. \ref{fig-setup-2}, where the coordinates
of the BS, IRS-1, IRS-2, U1, and U2 are the same as those in Fig.
\ref{fig-setup-1}. Fig. \ref{fig-SINRvsM-2I2U} plots the max-min
SINR as a function of $M$ for different choices of $N$, where we
set $d=5$m. Fig. \ref{fig-SINRvsN-2I2U} depicts the max-min SINR
as a function of $N$ for different values of $M$, where we set
$d=5$m. From Fig. \ref{fig-SINRvsM-2I2U} and Fig.
\ref{fig-SINRvsN-2I2U}, we see that the IRS-assisted system
outperforms the conventional massive MIMO system when $M\geq 260$,
and this advantage becomes more pronounced as the number of
reflect elements increases. Fig. \ref{fig-SINRvsM-2I2U} and Fig.
\ref{fig-SINRvsN-2I2U} also suggest that by increasing the number
of passive reflecting elements, one can achieve a same performance
with much fewer active antennas, therefore substantially reducing
the energy consumption at the BS. Fig. \ref{fig-SINRvsd-2I2U}
plots the max-min SINR as a function of $d$ for different values
of $N$, where we set $M=500$. From Fig. \ref{fig-SINRvsd-2I2U}, we
see that the max-min SINR improves substantially as users move
closer to IRSs, thus creating a signal hotspot in the vicinity of
IRSs.

\section{Conclusions}
\label{sec:conclusions} In this paper, we studied the problem of
joint active and passive beamforming for IRS-assisted massive MIMO
systems, where multiple IRSs equipped with a large number of
passive elements are deployed to assist the BS to simultaneously
serve a small number of single-antenna users. We aimed to maximize
the minimum SINR at users by jointly optimizing the transmit
precoding vector, the transmit power, and the phase shift
parameters. To address this problem, we first proved an important
and appealing property, referred to as AIC. The key idea is that
when an IRS is optimally tuned to serve a certain user, this IRS
will become interference-free to other users. By resorting to this
property, we came up with a simple yet effective solution to the
joint beamforming problem. Theoretical and simulation results
revealed that our proposed solution attains an SINR that scales
quadratically with the number of reflecting elements, and suggest
that, even with a moderate number of active antennas at the BS, a
massive MIMO like gain can be achieved via increasing the number
of passive reflecting elements.

\bibliography{newbib}

\begin{thebibliography}{10}
\providecommand{\url}[1]{#1}
\csname url@rmstyle\endcsname
\providecommand{\newblock}{\relax}
\providecommand{\bibinfo}[2]{#2}
\providecommand\BIBentrySTDinterwordspacing{\spaceskip=0pt\relax}
\providecommand\BIBentryALTinterwordstretchfactor{4}
\providecommand\BIBentryALTinterwordspacing{\spaceskip=\fontdimen2\font plus
\BIBentryALTinterwordstretchfactor\fontdimen3\font minus
  \fontdimen4\font\relax}
\providecommand\BIBforeignlanguage[2]{{%
\expandafter\ifx\csname l@#1\endcsname\relax
\typeout{** WARNING: IEEEtran.bst: No hyphenation pattern has been}%
\typeout{** loaded for the language `#1'. Using the pattern for}%
\typeout{** the default language instead.}%
\else
\language=\csname l@#1\endcsname
\fi
#2}}

\bibitem{RusekPersson13}
F.~Rusek, D.~Persson, B.~K. Lau, E.~G. Larsson, T.~L. Marzetta, O.~Edfors, and
  F.~Tufvesson, ``Scaling up {MIMO}: Opportunities and challenges with
  verylarge arrays,'' \emph{IEEE Signal Process. Mag.}, vol.~30, no.~1, pp.
  40--60, Jan. 2013.

\bibitem{LarssonEdfors14}
E.~G. Larsson, O.~Edfors, F.~Tufvesson, and T.~L. Marzetta, ``Massive {MIMO}
  for next generation wireless systems,'' \emph{IEEE Commun. Mag.}, vol.~52,
  no.~2, pp. 186--195, Feb. 2014.

\bibitem{Marzetta10}
T.~L. Marzetta, ``Noncooperative cellular wireless with unlimited numbers of
  base station antennas,'' \emph{IEEE Trans. Wireless Commun.}, vol.~9, no.~11,
  pp. 3590--3600, Nov. 2010.

\bibitem{BuzziI16}
S.~Buzzi, C.-L. I, T.~E. Klein, H.~V. Poor, C.~Yang, and A.~Zappone, ``A survey
  of energy-efficient techniques for 5{G} networks and challenges ahead,''
  \emph{IEEE J. Sel. Areas Commun.}, vol.~34, no.~4, pp. 697--709, April 2016.

\bibitem{ZhangWu17}
S.~Zhang, Q.~Wu, S.~Xu, and G.~Y. Li, ``Fundamental green tradeoffs:
  Progresses, challenges, and impacts on 5{G} networks,'' \emph{IEEE Commun.
  Surveys Tuts.}, vol.~19, no.~1, pp. 33--56, Firstquarter 2017.

\bibitem{TanSun18}
X.~Tan, Z.~Sun, D.~Koutsonikolas, and J.~M. Jornet, ``Enabling indoor mobile
  millimeter-wave networks based on smart reflect-arrays,'' in \emph{IEEE
  INFOCOM 2018 - IEEE Conference on Computer Communications}, Honolulu, HI,
  USA, April 16-19 2018, pp. 270--278.

\bibitem{AbariBharadia17}
O.~Abari, D.~Bharadia, A.~Duffield, and D.~Katabi, ``Enabling high quality
  untethered virtual reality,'' in \emph{14th USENIX Symposium on Networked
  Systems Design and Implementation (NSDI 17)}, Boston, MA: USENIX Association,
  March 27-29, 12-17 2017, pp. 531--544.

\bibitem{LiaskosNie18}
C.~Liaskos, S.~Nie, A.~Tsioliaridou, A.~Pitsillides, S.~Ioannidis, and
  I.~Akyildiz, ``A new wireless communication paradigm through
  software-controlled metasurfaces,'' \emph{IEEE Commun. Mag.}, vol.~56, no.~9,
  pp. 162--169, Sep. 2018.

\bibitem{HuRusek18}
S.~Hu, F.~Rusek, and O.~Edfors, ``Beyond massive {MIMO}: The potential of data
  transmission with large intelligent surfaces,'' \emph{IEEE Trans. Signal
  Process.}, vol.~66, no.~10, pp. 2746--2758, May 2018.

\bibitem{LiangLong19}
Y.-C. Liang, R.~Long, Q.~Zhang, J.~Chen, H.~V. Cheng, and H.~Guo, ``Large
  intelligent surface/antennas ({LISA}): Making reflective radios smart,''
  \emph{available at arXiv: 1906.06578}, 2019.

\bibitem{WuZhang19d}
Q.~Wu and R.~Zhang, ``Towards smart and reconfigurable environment: Intelligent
  reflecting surface aided wireless network,'' \emph{available at arXiv:
  1905.00152}, 2019.

\bibitem{HuangAlexandropoulos18}
C.~Huang, G.~C. Alexandropoulos, A.~Zappone, M.~Debbah, and C.~Yuen, ``Energy
  efficient multi-user {MISO} communication using low resolution large
  intelligent surfaces,'' in \emph{2018 IEEE Globecom Workshops (GC Wkshps)},
  Abu Dhabi, United Arab Emirates, Dec. 9-13 2018, pp. 1--6.

\bibitem{WuZhang19b}
Q.~Wu and R.~Zhang, ``Beamforming optimization for intelligent reflecting
  surface with discrete phase shifts,'' in \emph{ICASSP 2019 - 2019 IEEE
  International Conference on Acoustics, Speech and Signal Processing
  (ICASSP)}, Brighton, United Kingdom, May, 12-17 2019, pp. 7830--7833.

\bibitem{WuZhang18}
------, ``Intelligent reflecting surface enhanced wireless network: Joint
  active and passive beamforming design,'' in \emph{2018 IEEE Global
  Communications Conference (GLOBECOM)}, Abu Dhabi, United Arab Emirates, Oct.,
  9-13 2018, pp. 1--6.

\bibitem{WangFang19a}
P.~Wang, J.~Fang, X.~Yuan, Z.~Chen, H.~Duan, and H.~Li, ``Intelligent
  reflecting surface-assisted millimeter wave communications: Joint active and
  passive precoding design,'' \emph{available at arXiv: 1908.10734}, 2019.

\bibitem{NadeemKammoun19}
Q.-U.-A. Nadeem, A.~Kammoun, A.~Chaaban, M.~Debbah, and M.-S. Alouini,
  ``Asymptotic analysis of large intelligent surface assisted {MIMO}
  communication,'' \emph{available at arXiv: 1903.08127}, 2019.

\bibitem{WuZhang19a}
Q.~Wu and R.~Zhang, ``Intelligent reflecting surface enhanced wireless network
  via joint active and passive beamforming,'' \emph{IEEE Trans. Wireless
  Commun.}, vol.~18, no.~11, pp. 5394--5409, Nov. 2019.

\bibitem{HanTang19}
Y.~Han, W.~Tang, S.~Jin, C.-K. Wen, and X.~Ma, ``Large intelligent
  surface-assisted wireless communication exploiting statistical {CSI},''
  \emph{IEEE Trans. Veh. Technol.}, vol.~68, no.~8, pp. 8238--8242, Aug. 2019.

\bibitem{WuZhang19c}
Q.~Wu and R.~Zhang, ``Joint active and passive beamforming optimization for
  intelligent reflecting surface assisted {SWIPT} under {Q}o{S} constraints,''
  \emph{available at arXiv: 1910.06220}, 2019.

\bibitem{WangFang19b}
P.~Wang, J.~Fang, H.~Duan, and H.~Li, ``Compressed channel estimation and joint
  beamforming for intelligent reflecting surface-assisted millimeter wave
  systems,'' \emph{available at arXiv: 1911.07202}, 2019.

\bibitem{HeYuan19}
Z.-Q. He and X.~Yuan, ``Cascaded channel estimation for large intelligent
  metasurface assisted massive {MIMO},'' \emph{IEEE Wireless Commun. Lett.},
  pp. 1--1, 2019.

\bibitem{HuangZappone18}
C.~Huang, A.~Zappone, M.~Debbah, and C.~Yuen, ``Achievable rate maximization by
  passive intelligent mirrors,'' in \emph{2018 IEEE International Conference on
  Acoustics, Speech and Signal Processing (ICASSP)}, Calgary, AB, Canada, April
  15-20 2018, pp. 3714--3718.

\bibitem{YangZhang19}
Y.~Yang, S.~Zhang, and R.~Zhang, ``{IRS}-enhanced {OFDM}: Power allocation and
  passive array optimization,'' \emph{available at arXiv: 1905.00604}, 2019.

\bibitem{CuiZhang19}
M.~Cui, G.~Zhang, and R.~Zhang, ``Secure wireless communication via intelligent
  reflecting surface,'' \emph{IEEE Wireless Commun. Lett.}, vol.~8, no.~5, pp.
  1410--1414, Oct. 2019.

\bibitem{YuXu19}
X.~Yu, D.~Xu, and R.~Schober, ``Enabling secure wireless communications via
  intelligent reflecting surfaces,'' \emph{available at arXiv: 1904.09573},
  2019.

\bibitem{ShenXu19}
H.~Shen, W.~Xu, S.~Gong, Z.~He, and C.~Zhao, ``Secrecy rate maximization for
  intelligent reflecting surface assisted multi-antenna communications,''
  \emph{IEEE Commun. Lett.}, vol.~23, no.~9, pp. 1488--1492, Sep. 2019.

\bibitem{GuanWu19}
X.~Guan, Q.~Wu, and R.~Zhang, ``Intelligent reflecting surface assisted secrecy
  communication via joint beamforming and jamming,'' \emph{available at arXiv:
  1907.12839}, 2019.

\bibitem{MishraJohansson19}
D.~Mishra and H.~Johansson, ``Channel estimation and low-complexity beamforming
  design for passive intelligent surface assisted {MISO} wireless energy
  transfer,'' in \emph{ICASSP 2019 - 2019 IEEE International Conference on
  Acoustics, Speech and Signal Processing (ICASSP)}, Brighton, United Kingdom,
  May 12-17 2019, pp. 4659--4663.

\bibitem{Muhi-EldeenIvrissimtzis10}
Z.~Muhi-Eldeen, L.~Ivrissimtzis, and M.~Al-Nuaimi, ``Modelling and measurements
  of millimetre wavelength propagation in urban environments,'' \emph{IET
  Microwaves, Antennas Propag.}, vol.~4, no.~9, pp. 1300--1309, Sept. 2010.

\bibitem{CaiQuek11}
D.~W.~H. Cai, T.~Q.~S. Quek, and C.~W. Tan, ``A unified analysis of max-min
  weighted {SINR} for {MIMO} downlink system,'' \emph{IEEE Trans. Signal
  Process.}, vol.~59, no.~8, pp. 3850--3862, Aug. 2011.

\bibitem{TanChiang11}
C.~W. Tan, M.~Chiang, and R.~Srikant, ``Maximizing sum rate and minimizing mse
  on multiuser downlink: Optimality, fast algorithms and equivalence via
  max-min {SINR},'' \emph{IEEE Trans. Signal Process.}, vol.~59, no.~12, pp.
  6127--6143, Dec 2011.

\bibitem{HurKim13}
S.~Hur, T.~Kim, D.~J. Love, J.~V. Krogmeier, T.~A. Thomas, and A.~Ghosh,
  ``Millimeter wave beamforming for wireless backhaul and access in small cell
  networks,'' \emph{IEEE Trans. Commun.}, vol.~61, no.~10, pp. 4391--4403, Oct.
  2013.

\bibitem{ErcegGreenstein99}
V.~Erceg, L.~J. Greenstein, S.~Y. Tjandra, S.~R. Parkoff, A.~Gupta, B.~Kulic,
  A.~A. Julius, and R.~Bianchi, ``An empirically based path loss model for
  wireless channels in suburban environments,'' \emph{IEEE J. Sel. Areas
  Commun.}, vol.~17, no.~7, pp. 1205--1211, July 1999.

\end{thebibliography}
\bibliographystyle{IEEEtran}

\end{document}